\documentclass{article}
\usepackage[utf8]{inputenc}

\usepackage{amsmath,amsfonts,amssymb}
\usepackage{systeme}
\usepackage{hyperref}
\usepackage{graphicx}
\usepackage[natbib=true,style=apa]{biblatex}
\usepackage{float}
\usepackage{floatpag}

\usepackage[utf8]{inputenc} 
\usepackage[T1]{fontenc}    
\usepackage{url}            
\usepackage{booktabs}       
\usepackage{amsfonts}       
\usepackage{mathtools}
\usepackage{nicefrac}       
\usepackage{microtype}      
\usepackage{xcolor}         
\usepackage{times}
\usepackage[a4paper, total={5.5in, 8in}]{geometry}
\usepackage{subfiles}

\newcommand{\bracket}[3]{\left#1 #3 \right#2}

\renewcommand{\b}{\bracket{(}{)}}
\newcommand{\sqb}{\bracket{[}{]}}

\newcommand{\yh}{\hat{y}}

\DeclareMathOperator{\E}{E}

\bibliography{Bibliography}

\usepackage[section]{placeins} 

\title{What deep reinforcement learning tells us about human motor learning and vice-versa}
\author{Michele Garibbo\textsuperscript{1}, Casimir Ludwig\textsuperscript{2}, Nathan Lepora\textsuperscript{1} and Laurence Aitchison\textsuperscript{3}\\
\\
\textsuperscript{1}Department of Engineering Mathematics, Faculty of Engineering, University of Bristol, UK \\ \textsuperscript{2}School of Experimental Psychology, University of Bristol, UK \\\textsuperscript{3}Department of Computer Science, Faculty of Engineering, University of Bristol, UK}

\date{}

\begin{document}

\maketitle


Machine learning and specifically reinforcement learning (RL) has been extremely successful in helping us to understand neural decision making processes.
However, RL's role in understanding other neural processes especially motor learning is much less well explored.
To explore this connection, we investigated how recent deep RL methods correspond to the dominant motor learning framework in neuroscience, error-based learning.
Error-based learning can be probed using a mirror reversal adaptation paradigm, where it produces distinctive qualitative predictions that are observed in humans.
We therefore tested three major families of modern deep RL algorithm on a mirror reversal perturbation.
Surprisingly, all of the algorithms failed to mimic human behaviour and indeed displayed qualitatively different behaviour from that predicted by error-based learning.
To fill this gap, we introduce a novel deep RL algorithm: model-based deterministic policy gradients (MB-DPG).
MB-DPG draws inspiration from error-based learning by explicitly relying on the \textit{observed} outcome of actions.
We show MB-DPG captures (human) error-based learning under mirror-reversal and rotational perturbation.
Next, we demonstrate error-based learning in the form of MB-DPG learns faster than canonical model-free algorithms on complex arm-based reaching tasks, while being more robust to (forward) model misspecification than model-based RL.
These findings highlight the gap between current deep RL methods and human motor adaptation and offer a route to closing this gap, facilitating future beneficial interaction between the two fields.


\section{Introduction}

The relationship between (artificial) reinforcement learning, RL, and neuroscience has formed a "virtuous cycle", with neuroscience informing RL, and RL informing neuroscience \citep{hassabis2017neuroscience, botvinick2020deep}.
For instance temporal difference algorithms, among the most powerful methods in RL, have been developed in close interaction with neuroscience and psychology \citep{sutton1987temporal,sutton1990time, watkins1992q}.
In turn, neuroscience has benefited from using principled RL methods to describe neural data (e.g.\ dopamine signals) and behavioural data (e.g. choices) \citep{barto1995adaptive,montague1996framework,schultz1997neural,bayer2005midbrain,niv2009reinforcement}. 
Recent dramatic advances in deep RL raise the prospect of understanding complex, realistic animal behaviours \citep[see][for a review]{botvinick2020deep}.

This virtuous cycle between neuroscience and RL has primarily focused on reward-based learning in canonical decision-making tasks \citep[e.g.][]{niv2009reinforcement,botvinick2020deep}.
In this setting, action choices are typically associated with a reward signal or value (i.e. long-term reward), encoding the degree of success or failure \citep[][]{wolpert2011principles, niv2009reinforcement}.
Optimal actions are then learned by choosing those actions associated with the highest reward.
This process has primarily been linked to dopamine signals in the ventral tegmental area, which are thought to encode reward prediction errors  \citep[]{luft2009dopaminergic,hosp2011dopaminergic,izawa2011learning,schultz1997neural,niv2009reinforcement}.

However, reward-based learning does not appear to be the exclusive learning mechanism underlying  action and movement --here, error-based learning mechanisms seem to be key \citep[][]{wolpert2011principles,izawa2011learning}. 
Error-based learning is distinct from reward-based learning because it uses richer information capturing the relation (gradient) between actions and their sensory outcomes \citep[these gradients capture ``directed errors'', conveying the direction in which actions should be changed to correct for a perceived sensory error - i.e.  action gradients; ][]{wolpert2011principles}.
The distinction between error- and reward- based learning is also reflected in the underlying neural circuitry, with error-based learning being associated with complex spikes in cerebellar Purkinje cells \citep[][]{kitazawa1998cerebellar,ito2013error, dean2010cerebellar,tseng2007sensory}.
In the motor neuroscience literature, error-based learning is typically probed with adaptation paradigms \citep{synofzik2006internalizing, mazzoni2006implicit, miall1996forward,izawa2011learning} \citep[see also][for a review]{krakauer2019motor}.
In these paradigms, a perturbation is introduced in the sensory feedback following an action.
For example, the visual endpoint of a reaching movement may be displaced from its true location.
Initially such a perturbation introduces an error in the action, which is typically corrected over time, reflecting error-based learning mechanisms.
That is, suppose the visual endpoint is displaced to the left of the target in a reaching task. 
The gradient information of this visual error would specify a more rightward reach to correct the error, whereas a reward signal might merely convey the distance to the target with no directional information.

In the past few years, major advances have been made towards producing deep RL algorithms that can control simulated physical bodies.
These physics-based control tasks involve learning a controller (i.e.\ policy) capable of actuating a simulated or real robotic ``body'', achieving a new horizon of motor control within complex naturalistic settings \citep[e.g.][]{heess2016learning, merel2018hierarchical, schulman2017proximal,akkaya2019solving,silver2014deterministic}.
As such, these advances in deep RL might help us to understand human motor learning, especially in highly complex, naturalistic settings ,\citep[e.g.][]{fischer2021reinforcement}.
However, we do not know how these methods relate to the dominant motor learning framework in neuroscience, error-based learning and, indeed, whether they capture basic phenomena from motor neuroscience.
To this end, we aim to elucidate the relation between deep RL methods and error-based learning.
As a first step, we considered the qualitative behaviour of deep RL algorithms under a mirror reversal perturbation of visual feedback during arm reaches.
In humans, mirror reversal perturbations cause an increase in error adaptation as learning progresses, providing a distinctive learning prediction to characterise error-based learning \citep[as shown across several studies - e.g. ][]{hadjiosif2021did,lillicrap2013adapting,wang2021implicit,krakauer2019motor,kasuga2015alteration}.
We tested three major families of RL algorithms under a mirror reversal task, and found that they did not replicate this (human) learning prediction.
This failure was surprising, because at least two of these algorithms leverage gradients of estimated rewards with respect to actions (i.e. compute "directed errors" in the form of action gradients), and we therefore expected that they would resemble error-based learning.

To address this fundamental difference between modern deep RL and human error-based learning, we introduce an alternative deep RL algorithm, model-based deterministic policy gradient (MB-DPG).
MB-DPG draws inspiration from (human) error-based learning, bringing computational mechanisms from the neuroscience literature \citep[i.e. forward model differentiation - ][]{jordan1992forward} to the deep RL setting.
We found MB-DPG accurately reproduces human experimental data in mirror-reversal tasks.
We then tested how MB-DPG behaved in an adaptation paradigm where a rotational perturbation was applied to the movement.
Under error-based learning a perturbation affecting the \textit{direction} of the action gradient (e.g.\ mirror-reversal) should be more disruptive than a perturbation affecting  its \textit{magnitude} (e.g.\ rotation), due to the reliance on "directed" error signals  \citep[see][]{abdelghani2008sensitivity}.
Indeed, humans adapt more easily to such rotational perturbations and we found that MB-DPG recovers more quickly as well compared to the mirror reversal \citep[][]{lillicrap2013adapting}.

We then introduced a more complex simulated arm reaching task \citep[][]{berret_chiovetto_nori_pozzo_2011}, going beyond adaptation paradigms, and assessing MB-DPG ability to support motor learning completely from scratch. 
We found that MB-DPG learned faster than canonical model-free policy gradient methods such as REINFORCE \citep[see][]{williams1992simple} and DDPG \citep[see][]{lillicrap2015continuous, silver2014deterministic}. 
Additionally, a critical aspect of human motor learning and adaptation is its robustness under rapid changes in bodily state (e.g. through tiredness or injury) \citep{wolpert2001perspectives,shadmehr2010error}.
Such rapid changes result in a mis-specified forward model: that is, the forward model no longer reflects the (new) bodily state.
Thus, any motor adaptation strategy used by humans must be robust to this mis-specification.
Indeed, we found that MB-DPG was more robust to mis-specified forward models than pure model-based methods on human-like reaching tasks, as it used observed rather than predicted outcomes to drive adaptation.

\section{Results}

\subsection{Canonical deep RL methods fail to mimic human error-based learning}

\begin{figure} 
\centering
\includegraphics[width=150mm,scale=1.5]{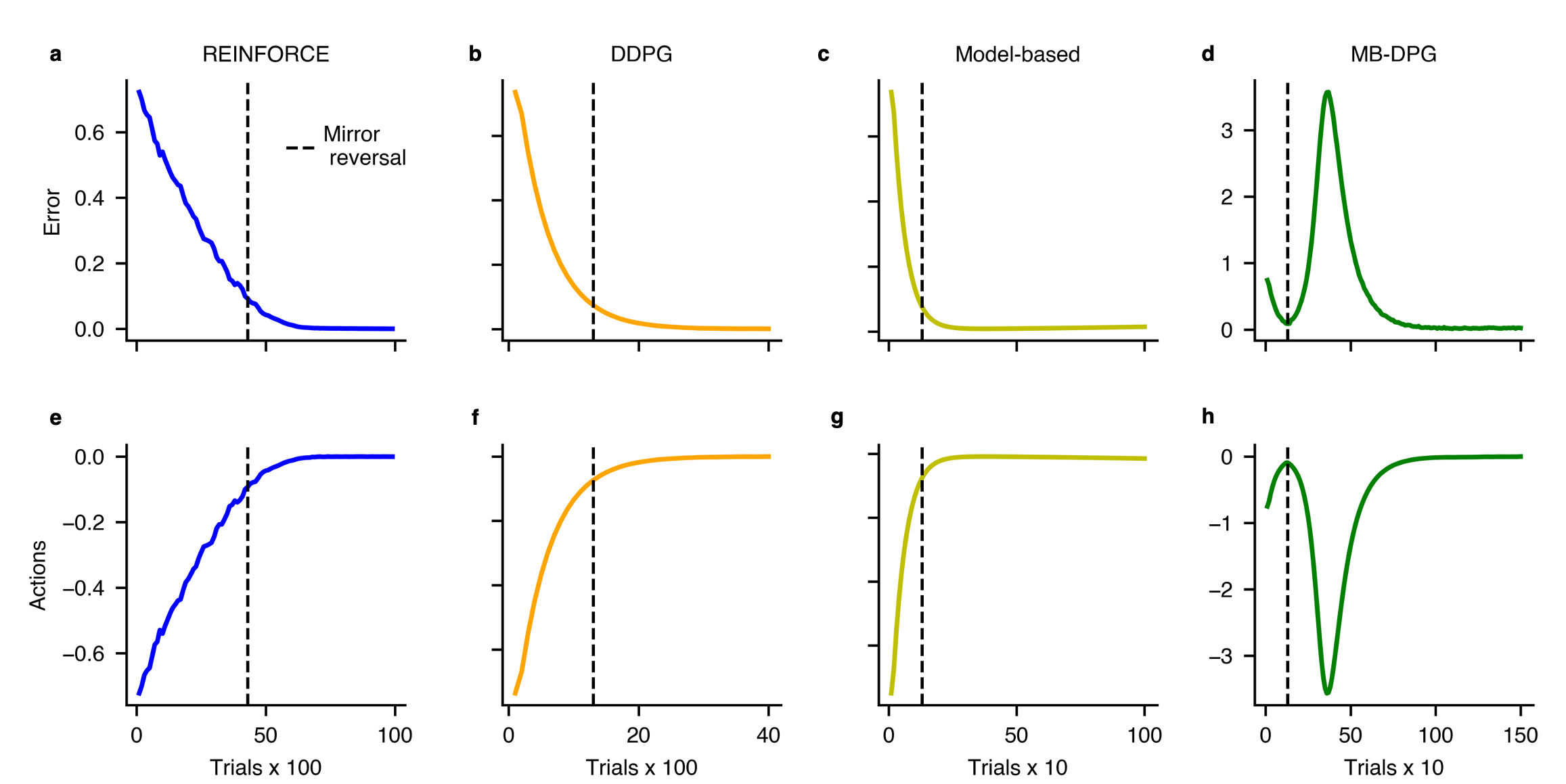}
\caption{Accuracy (top row) and motor control signal (action) (bottom row) of each algorithm in the mirror-reversal task. The vertical dotted line denotes the point at which the mirror-reversal perturbation was introduced.}
\label{fig:MR_all_result}
\end{figure}

In this section, we tested whether canonical deep RL methods can reproduce human error-based learning under the mirror-reversal perturbation of targeted arm reaches.
As discussed in the introduction, this perturbation allows to have a specific prediction on learning performance under error-based learning \citep[i.e. performance should get worse as learning progresses, ][]{hadjiosif2021did,lillicrap2013adapting,wang2021implicit,krakauer2019motor,kasuga2015alteration}).
This specific prediction provides a qualitative measure to determine potential differences between (human) error-based learning and deep RL methods, which would otherwise be hard to assess under normal settings.
We tested 3 major families of (deep) RL algorithms on the mirror reversal task: a canonical model-free policy gradient method, REINFORCE \citep[][]{williams1992simple}, a model-free value-based method, DDPG \citep[][]{lillicrap2015continuous, silver2014deterministic}, and finally, a model-based policy gradient method \citep[][]{clavera2020model,heess2016learning}.
We found that all tested deep RL algorithms were unaffected by the mirror-reversal perturbation (see Fig.~\ref{fig:MR_all_result}a-c), in the sense that performance continued to improve immediately after the perturbation was introduced (dashed vertical line).
These results contrasted error-based learning observed in humans, in which performance worsens immediately after the perturbation  \citep[e.g.][]{hadjiosif2021did,lillicrap2013adapting,wang2021implicit} (see also Fig.~\ref{fig:HumanLike_MBDPG}a).

\begin{figure} 
\centering
\includegraphics[width=150mm]{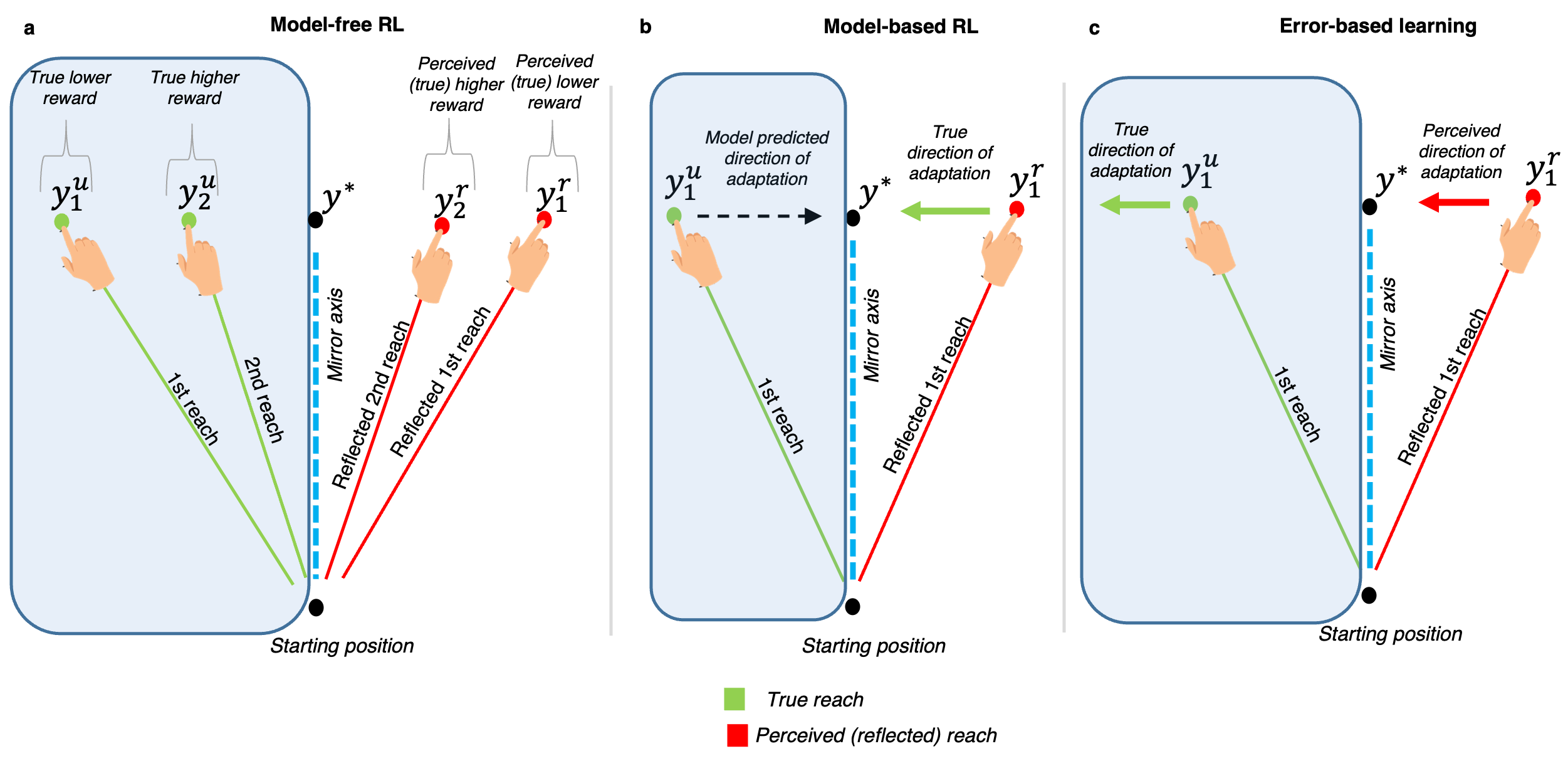}
\caption{ Error-based learning (humans) (\textbf{a}) Model-free RL (\textbf{b}) and Model-based RL (\textbf{c}) under mirror-reversal. The position of the target is indicated by $y^*$ ; $y^u$ indicates the true (non-reflected) outcome of a reaching attempt, which is hidden (here represented by the shaded region). Only the mirror-reflected outcome, denoted as $y^\text{r}$, can be perceived, inducing the illusion of having reached on the opposite (mirrored) side of the target.
\textbf{a}. In error-based learning, the direction of adaptation is determined by where the arm is perceived to be and thus, is inverted under mirror-reversal. 
That is, the perceived direction of adaptation needed to reach $y^*$ from $y_1^r$ (horizontal red line) correspond to the wrong direction of adaption needed to reach $y^*$ from the true reaching outcome, $y_1^u$ (i.e. moving away from the target). 
Thus, under error-based learning, performance gets worse when mirror-reversal is introduced.
\textbf{b}. In model-free, the reward structure is preserved under mirror-reversal. 
That is, reaches perceived (red lines) as more rewarding (i.e. closer to $y^*$) under the reflection correspond to the true (green lines) more rewarding reaches, implying performance should not be affected by mirror-reversal (i.e. still moving towards the target by prioritising high reward reaches).
\textbf{c}. In model-based RL, the direction of adaptation is determined by where the model predicts the arm to be (i.e. dashed black line).
In this case, the model still predicts the arm to be in the true non-mirror position (i.e. left side of the target), nullifying the mirror-reversal effect of perceiving the arm on the wrong side of the target.
This allows to adapt reaches in the correct direction, explaining why performance is unaffected by the mirror perturbation.}
\label{fig:MR_RL_vs_Er_learning}
\end{figure}

\begin{figure} 
\centering
\includegraphics{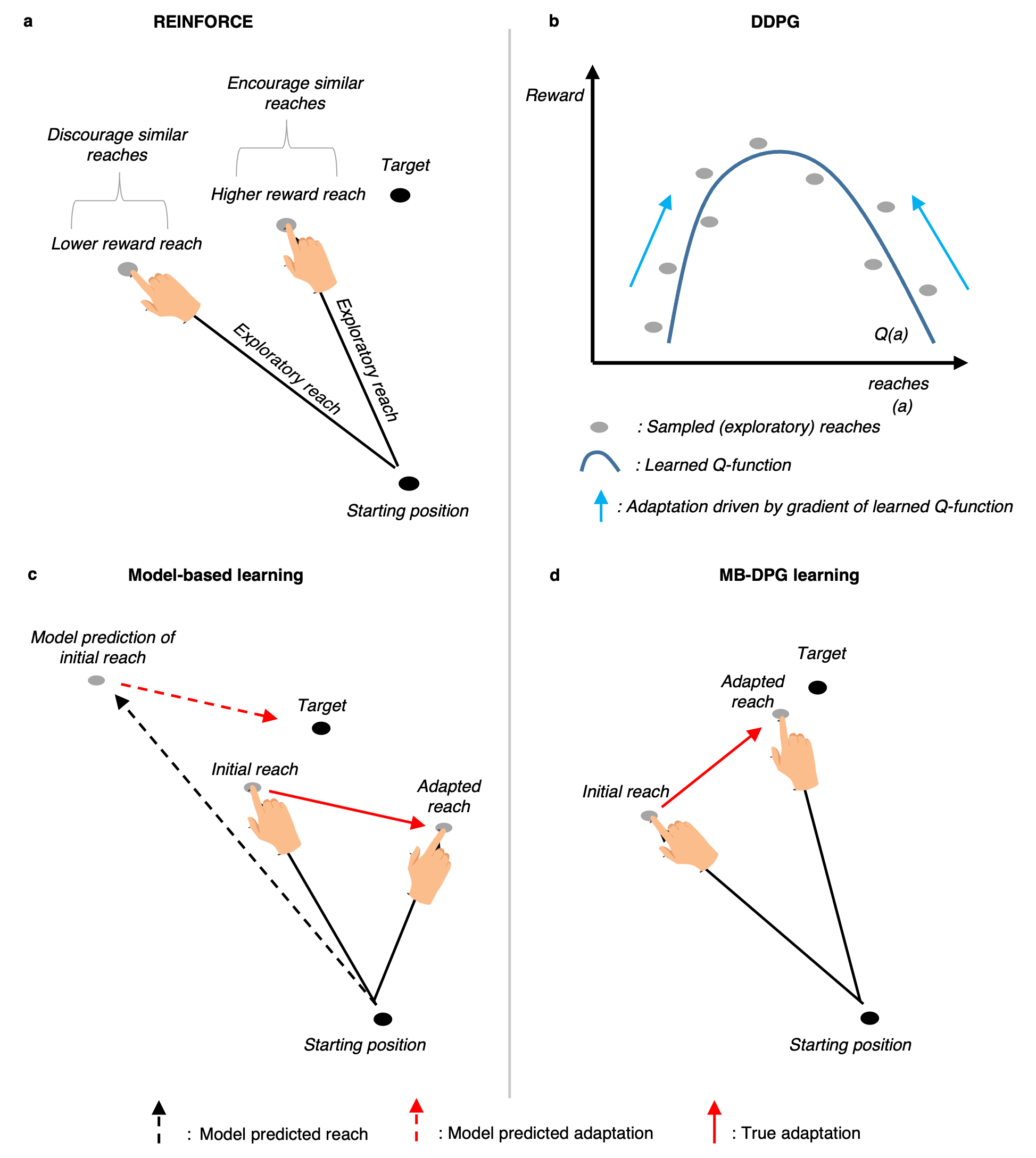}
\caption{Diagram of how the four tested algorithms learn. \textbf{a}. REINFORCE learns by trying exploratory reaches, prioritising those reaches that get closer to the target, while discouraging those further away from the target. This process is iterated until the target is reached. \textbf{b}. DDPG employs exploratory reaches to learn a differentiable reward function, Q-function, which instructs how to change reaches to gain more reward (blue solid line). Crucially, this Q-function is based on the action space and thus, does not depend on where reaches land in space relative to the target (e.g. to the left or to the right of the target). \textbf{c}. Model-based learns based on the model predicted outcome, by computing the error from the model prediction to the target (dotted red line) and using this error to change the initial reach (solid red line). In the illustration, the model prediction is wrong since it does not correspond to the initial reach, resulting in incorrect adaptation (i.e. the adapted reach is further away from the target than the initial reach). \textbf{d}. MB-DPG learns based on the observed outcome, by computing the error signal from the outcome of the (initial) reach to the target, thus enabling accurate adaption (i.e. the adapted reach is closer to the target).}
\label{fig:Learning_diagram}
\floatpagestyle{empty}
\end{figure}

Why do each of these algorithms fail to reproduce human behaviour?
First, REINFORCE chooses random actions, then reinforces those that lead to higher reward than expected (Fig.~\ref{fig:Learning_diagram}a; Methods \ref{algorithms}).
Critically, mirror reversal changes only the observed outcomes (position of the hand), but does not change the mapping from actions to rewards.
As REINFORCE updates depend only on the actions and associated rewards (i.e. model-free), mirroring reaching outcomes has no effect, as highlighted in Fig. (\ref{fig:MR_RL_vs_Er_learning}b).
Second, DDPG trains a neural network to predict the reward from the action, then does gradient ascent to find actions and policies with higher predicted rewards (Fig.~\ref{fig:Learning_diagram}b; Methods \ref{algorithms}).
Critically, this process depends only on the actions and rewards, and is again independent of the observed outcome (i.e. DDPG is also model-free).
As it is only the observed outcome that is perturbed under mirror reversal, DDPG is unaffected by mirror reversal (i.e. similarly to REINFORCE - Fig. (\ref{fig:MR_RL_vs_Er_learning}b).
Finally, model-based RL explicitly learns a model mapping from action to outcomes, then predicts rewards from the outcome (using a learned or known function mapping outcomes to rewards). 
Then, model-based RL does gradient ascent to find actions or policies with higher rewards (Methods \ref{algorithms}).
Critically, the model is learned using actions and the \textit{observed} outcome, so there is at least some outcome dependence.
However, while the model updates depend on outcomes, individual updates to the actions do not.
In fact, an individual update to the action depends only on the (model) predicted rewards: with the action/policy changing slightly to improve \textit{predicted} reward based on the already-learned model (see Fig.~\ref{fig:Learning_diagram}c).
Thus, as the mirror-reversal is introduced, the learned model mapping actions to outcomes continues to predict that the reaching outcome will be in the true (non-mirrored) location, so reaches will be adapted in the right direction and the error continues to decrease immediately after mirror reversal (see Fig. (\ref{fig:MR_RL_vs_Er_learning}c for a visual explanation).
We find that the error continues to decrease, even as the model is updated with the mirror-reversed outcomes.
Conversely, human error-based learning appears to rely on the \textit{observed} reaching outcome to determine the necessary direction of changes to motor commands (e.g. observing a reach with a leftward error leads to a rightward correction on the next attempt).
As a result, when the observed sensory outcome is mirror-reflected, motor commands are changed in the wrong direction (see Fig. (\ref{fig:MR_RL_vs_Er_learning}a for a visual explanation), explaining the worsening of performance documented in humans under mirror-reversal  \citep[e.g.][]{hadjiosif2021did,lillicrap2013adapting,wang2021implicit}.
Therefore, unlike deep RL methods, (human) error-based learning appears to be driven by the \textit{observed} outcome of movements.

\subsection{MB-DPG: a deep RL algorithm that accounts for mirror reversal}
\label{MBDPG}

To account for the failure of canonical deep RL algorithms to reproduce human error-based learning, we consider an alternative deep RL algorithm, MB-DPG. 
By drawing inspiration from human error-based learning, MB-DPG explicitly learns based on the observed rather than the model predicted reaching outcome (see \ref{fig:Learning_diagram}d for a visual explanation).
To understand the computational basis of this property of human error-based learning, consider the relationship between rewards, (sensory) outcomes and action parameters, expressed as the product of three derivatives (shown for the 1d case for simplicity): 
\begin{align}\label{error_grad}
  \frac{dr}{d\phi} &= \frac{dr}{dy} \frac{dy}{da} \frac{da}{d\phi} 
\end{align}
where $r$ is the reward, $y$ is the outcome (i.e.\ the final position of the hand), $a$ is the action and $\phi$ is the policy parameters used to generate the action (e.g. synaptic strength).
This expression explicitly updates the policy parameters, $\phi$, using information about the mapping from actions to outcomes, $dy/da$ and the relationship between outcomes and rewards, $dr/dy$.
We can think of this outcome-to-reward relation, $dr/dy$, as a directed error signal, indicating the direction of movement to reach closer to the target on subsequent trials (i.e.\ to increase the reward).
However, $dy/da$ is the gradient of the actual outcome with respect to the action, and this cannot be known exactly, especially in complex settings, where our our actions are part of a complex and tightly coordinated sequence of motor commands.
Hence, the action-to-outcome relationship, $\frac{dy}{da}$, must be learned from experience \citep[][]{abdelghani2008sensitivity}. 
This can be done by learning a model of the dynamics, relating actions to their estimated outcomes, which we denote as $\hat{y}$ (Methods \ref{modelLearning}) \citep[e.g.][]{jordan1992forward}. 
Based on this learned model, we can approximate the gradient in (\ref{error_grad}) as:
\begin{align}\label{MBDPG_grad}
    &\text{MB-DPG:} & \frac{dr}{d\phi} &= \frac{dr}{dy} \frac{dy}{da} \frac{da}{d\phi} \approx \frac{dr}{dy} \frac{d \hat{y}}{da} \frac{da}{d\phi} 
\end{align}
where $\frac{d\yh}{da}$ represents the estimate of $\frac{dy}{da}$ in (\ref{error_grad}), computed by differentiating through the learned model of the dynamics.
We integrate this gradient computation within a deep RL algorithm, which we denote as MB-DPG.
Crucially, If we test this algorithm on the mirror-reversal perturbation, we see that when the perturbation is introduced (dashed vertical lines), MB-DPG performance starts to worsen as documented in humans \citep[e.g.][]{hadjiosif2021did,lillicrap2013adapting,wang2021implicit} and unlike canonical RL algorithms.
Eq.~\ref{MBDPG_grad} highlights the key aspect that drives this performance in MB-DPG (and humans): MB-DPG uses the gradient of the reward, $dr/dy$, at the actual final hand position, $y$, which 
ensures learning of the policy parameters, $\phi$, is driven by the observed outcome of actions.

This aspect of MB-DPG is fundamentally different from the model-based RL algorithm.
In model-based RL, the gradient of the reward is computed wrt the predicted outcome, $\hat{y}$,
\begin{align}\label{modelBased_grad}
    &\text{Model-based} & \frac{dr}{d\phi} &= \frac{dr}{dy} \frac{dy}{da} \frac{da}{d\phi} \approx \underbrace{\frac{dr}{d\hat{y}} \frac{d \hat{y}}{da}}_{\mathclap{{\text{model-predicted } dr/da}}} \frac{da}{d\phi} 
\end{align}
This is by far the most natural approach in a machine learning setting, as we can simply backpropagate through the model, from rewards to actions, to obtain $dr/da$, and this is something that modern deep learning frameworks are easily able to do \citep[i.e. denoted as SVG in the RL literature,][]{clavera2020model, heess2016learning}.
In contrast, combining $dr/dy$ and $d\hat{y}/da$ is much more difficult, requiring us to explicitly compute and manipulate these gradients separately.
While the model-based approach is easier to implement, it comes with a considerable disadvantage: its action update,$\frac{dr}{d\phi}$, is entirely independent of the observed outcome, $y$, and depends only on the model predicted outcome $\hat{y}$.
This causes the failure of model-based deep-RL methods to reproduce human behaviour on the mirror reversal task, and as we show below, considerably reduces robustness to misspecified forward models.
Finally, model-free RL algorithms also computationally differ from MB-DPG and human error-based learning.
This is because these algorithms ignore the explicit relation between rewards, $r$, and outcomes, $y$, by directly modelling the relationship between actions,$a$ and rewards:

\begin{align}\label{modelBased_grad}
    &\text{Model-free} & \frac{dr}{d\phi} &= \frac{dr}{da} \frac{da}{d\phi}
\end{align}

\noindent Therefore, a mirror-reversal of the outcome, $y$, does not seem to have any impact on these methods (i.e. REINFORCE and DDPG) as explained in the previous section.


Note that our proposed distinction between MB-DPG and model-based is consistent with the computational distinction proposed by \citet[][]{jordan1992forward} in terms of "predicted performance error" (i.e. analogous to Model-based) and "performance error" (i.e. analogous to MB-DPG).
We embed these error computations within modern deep RL framework, augmenting them with key RL mechanisms (e.g. exploration, memory reply - see Method \ref{algorithms}) thus, enabling comparisons between canonical deep RL methods and human performance.


\subsection{MB-DPG and human performance}

\begin{figure} 
\hspace*{-1cm}
\includegraphics[width=150mm]{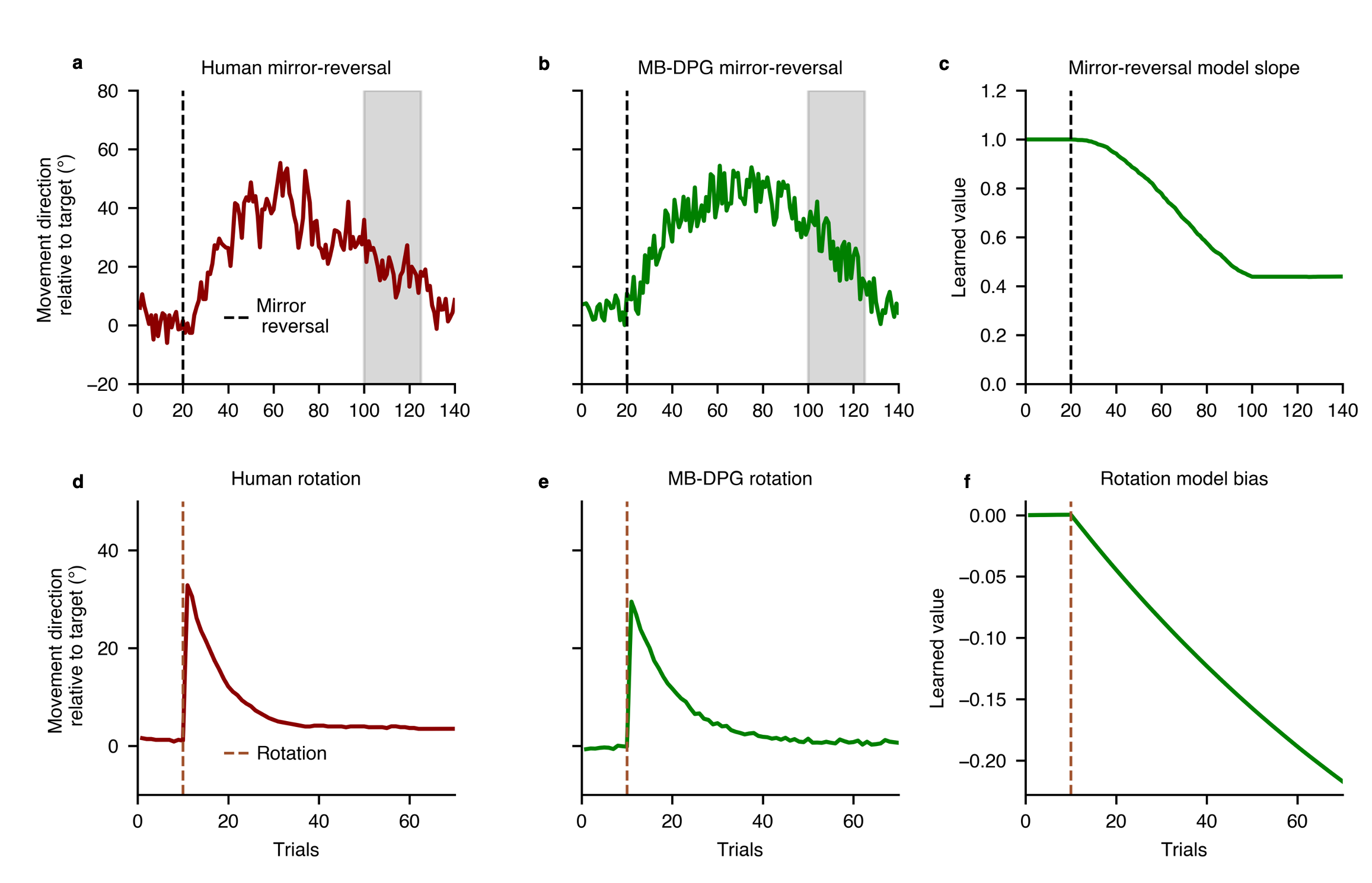}
\caption{Comparison between human performance and MB-DPG on the mirror-reversal (top row) and the rotation tasks (bottom row); In \textbf{a}. and \textbf{b}. the gray shade denotes the non-visual feedback phase, where the visual error is removed, leaving  proprioceptive feedback only; the final 15 trials after this phase represent the washout phase, in which the perturbation is removed. Human performance is shown in the mirror-reversal, \textbf{a}, and rotation, \textbf{d}, tasks averaged across participants, data respectively plotted from \citep{hadjiosif2021did} and \citep{krakauer2000learning} using WebPlotDigitizer \citep[][]{Rohatgi2020}. MB-DPG performance is shown in the mirror-reversal, \textbf{b}, and  rotation \textbf{e} tasks, using the update saturation and non-visual feedback. \textbf{c}. Learning of the (forward) model parameter during the mirror-reversal task; the mirror-reversal perturbation is learnt when this parameter becomes negative, capturing the (mirror-) reversal in the action-to-outcome relation. \textbf{f}. Learning of the (forward) model parameter during the rotation task; the rotation perturbation is learnt when this parameter reaches the same value as the magnitude of the rotation perturbation (i.e. 0.53 for a 30 degree visuomotor rotation).}
\label{fig:HumanLike_MBDPG}
\end{figure}

Fig~(\ref{fig:MR_all_result}) provides an initial test of how the different RL algorithms handle the mirror-reversal: can they capture the worsening of performance after the perturbation is introduced? We now turn to a more rigorous assessment of whether and to what extent MB-DPG can capture human performance.
First, in Fig.~\ref{fig:MR_all_result}d, MB-DPG appears to recover from the perturbation, allowing performance to finally converge. 
This is because MB-DPG eventually learns the inverted (mirrored) action-to-outcome relation (i.e. learns to invert the sign of $\frac{d \hat{y}}{da}$), enabling adaptation to occur in the right direction.
However, it is not clear how quickly humans can learn the inverted action-to-outcome relation, especially, for reaching movements, where very strong priors may exist (e.g. for our entire life, we experience that a leftward correction is needed for a rightward error) \citep[][]{hadjiosif2021did,lillicrap2013adapting,wang2021implicit}.
To capture this effect, we dramatically reduced the learning rate for the model.
Second, human performance seems to saturate after the initial worsening of performance triggered by the mirror-reversal \citep[see Fig.~\ref{fig:HumanLike_MBDPG}a, data plotted from][]{hadjiosif2021did}. 
The accuracy saturation seems to indicate that when the perturbation becomes too extreme, humans give less weight to visual sensory errors \citep[e.g.][]{krakauer2019motor,marko2012sensitivity,wei2009relevance}, reducing learning.
To capture this effect with MB-DPG, we saturate the MB-DPG error update as sensory errors become larger and larger (see Methods \ref{human_MBDPG}).
Third, humans return to baseline and thereby improve their performance in trials without any visual feedback (see gray shade trials in Fig.~\ref{fig:HumanLike_MBDPG}a), indicating that there must be additional (e.g.\ proprioceptive) sources of error information \citep[][]{hadjiosif2021did}.
After accounting for these three effects, MB-DPG is able to replicate the learning trends seen in humans under mirror-reversal (see Fig.~\ref{fig:HumanLike_MBDPG}b).
Similarly to humans, MB-DPG performance recovers slightly, before the visual feedback is withheld and they have to rely on non-visual feedback (gray shade).
This arises out of a complex interaction between a small amount of model learning (which reduces the error, but does not flip its direction), saturation and non-visual feedback.

Next, we tested MB-DPG against another human experimental prediction. 
Under error-based learning, perturbations that alter, but do not reverse the action-to-outcome relation (e.g. visuomotor rotations), should only affect learning marginally compared to perturbations that reverse the action-to-outcome relation (e.g. mirror-reversal) \citep[see ][for an in-depth explanation]{abdelghani2008sensitivity}.
This prediction is evident in human data, where on average, only a few trials are needed to recover from a fairly large 30 degree visuomotor rotation \citep[see Fig.~\ref{fig:HumanLike_MBDPG}d, data plotted from][]{krakauer2000learning}, while the same is not true for mirror-reversal (see Fig.~\ref{fig:HumanLike_MBDPG}a) \citep[see also][]{lillicrap2013adapting,wang2021implicit}.
The exact same phenomena can be observed under MB-DPG (i.e. Fig.~\ref{fig:HumanLike_MBDPG}b vs Fig.~\ref{fig:HumanLike_MBDPG}e).
By looking at MB-DPG, we can see this learning pattern can be explained by considering how the estimated (forward) model parameters affect learning in relation the perturbations.
Under the rotation, learning immediately improves performance, despite very inaccurate model estimates (see Fig.~\ref{fig:HumanLike_MBDPG}f, immediately after the rotation, the model is far from having learnt the perturbation), thus, allowing MB-DPG (and humans) to correct for the perturbation quickly before the model has adapted.
Conversely, under mirror-reversal, learning cannot improve performance until the model parameters have switched sign (i.e. in Fig.~\ref{fig:HumanLike_MBDPG}c the model parameter remains positive), requiring more time before being able to correct for the perturbation (Fig.~\ref{fig:MR_all_result}d), with the risk of not being able to correct for the perturbation, if the model's priors are too strong.

\subsection{MB-DPG and model-based algorithms learn faster than model-free methods on more complex arm-based reaching tasks}

Next, we assessed the ability of MB-DPG to learn more complex and realistic motor control tasks.
To do this, we employed a simulated biomechanical model of the human arm from the neuroscience literature \citep[][]{berret_chiovetto_nori_pozzo_2011}, testing how well MB-DPG could train this arm to reach for target locations with a ballistic movement (i.e. feedforward control). 
As a reference measure, we compared MB-DPG performance with the other canonical deep RL methods as well as giving human average reaching accuracy \citep[based on][]{van2004role}.
Note for humans, it is impossible to have a learning curve for this kind of task, since we learn to reach over years of practice during childhood. 
Each algorithm's performance was tested on two sub-tasks: a simple single-target and a more complex multi-target reaching task. 
In the single target setting, the arm had to learn to reach towards a single target location, whereas in the multi-target arm-based task, the model arm had to learn to reach towards 50 different (random) target locations within the same training run (see figure \ref{fig:Arm_diagram} a).

We found that model-based and MB-DPG performed similarly on both tasks, and they both performed better than REINFORCE and DDPG.
Moreover, the difference in performance was much bigger in the more complex multi-target task, with model-based and MB-DPG performing considerably better than REINFORCE and DDPG.
Note that we performed an extensive hyperparameter search for all algorithms (Methods \ref{hyperSearch}), and these results do not stem from the model-based methods merely being better tuned.
We noticed DDPG was very unstable, with good performance with some random seeds and bad performance with others. 
Indeed, the issue was so bad that the hyperparameter search gave parameter values that worked well with one set of seeds, but broke down when we changed the random seeds for our final run assessing performance, explaining DDPG poor performance.
This finding is consistent with previous claims of the high sensitivity of this algorithm to hyper-parameter values \citep[][]{matheron2019problem}. 
Overall, these findings indicate that either model-based or MB-DPG methods are indeed far superior to model-free learning when learning complex motor control tasks, and highlight the importance of "signed" error signals (error-based learning) over simple rewards in motor control.

It should be no surprise that MB-DPG and the Model-based RL algorithms performed similarly in this idealised setting. 
In particular, as the arm dynamics did not change over training, it was possible for both methods to learn an extremely accurate model mapping actions to outcomes.
In that case, the predicted outcome used by model-based is very similar to the actual outcome used by MB-DPG, resulting in very similar performance.
This raises the question of how the two algorithms perform in the absence of this idealised setting, such as when the learn model dynamics are increasingly wrong.

\subsection{MB-DPG is more robust to model errors than standard model-based methods}

The previous reaching task was highly idealised, since we assumed that the mapping between actions and outcomes could be learned precisely.
However, in practice, human models are likely to be highly approximate, either due to biases in estimation of sensory outcomes, or due to changes in the mapping from actions to outcomes, e.g.\ due to fatigue, injury, body change \citep[][]{rieser1995calibration}.
Here, we sought to understand the implications of these model errors on the performance of Model-based RL and MB-DPG algorithms.

\begin{figure} 
\centering
\includegraphics[width=150mm]{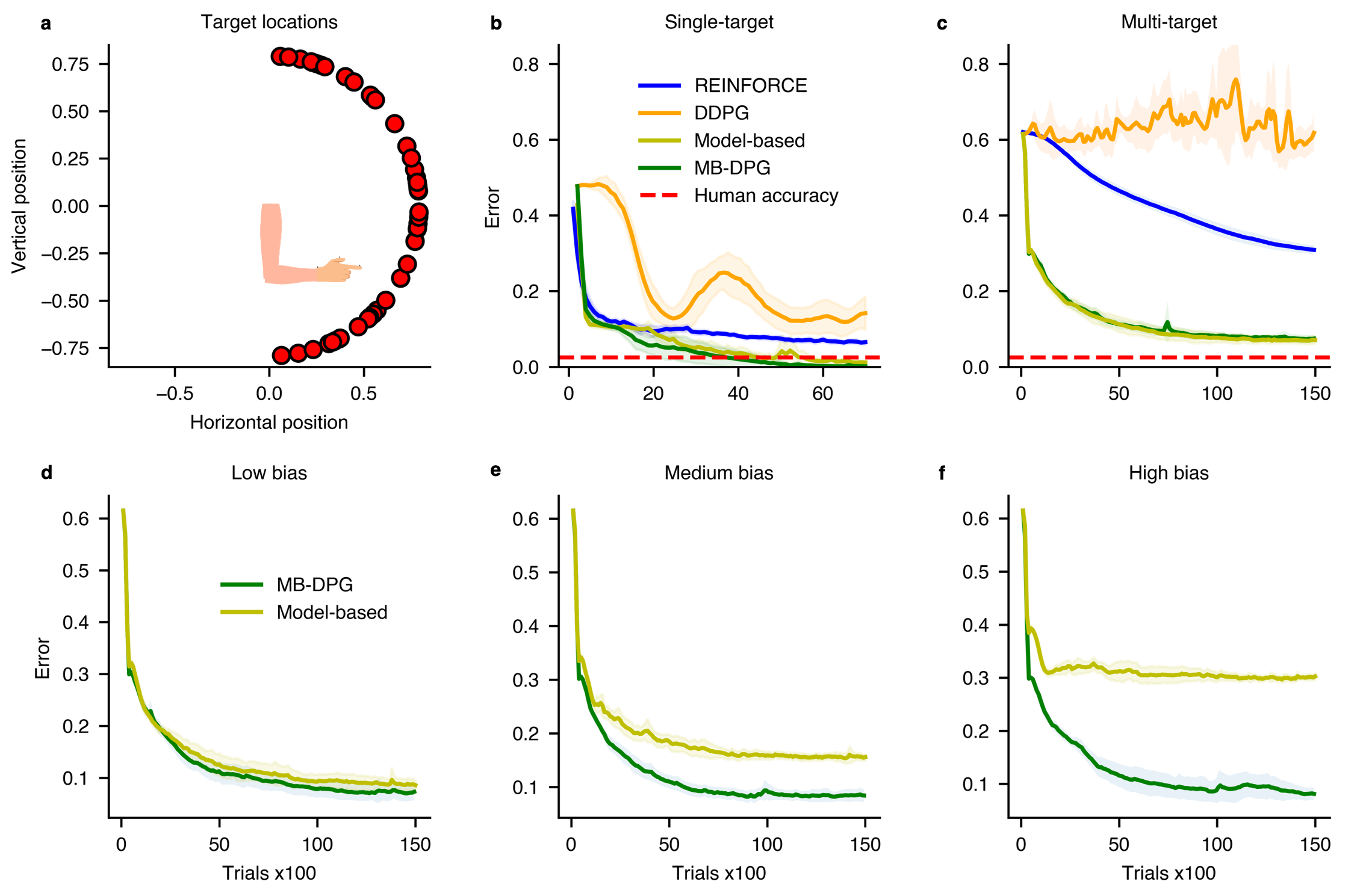}
\caption{Diagram and results for the arm-based reaching tasks. \textbf{a}. Arm representation in relation to the 50 target locations in the multi-target reaching task on a 2 dimensional horizontal plane (the algorithms had to learn to reach towards each location through one unique policy). \textbf{b}. Accuracy comparison across algorithms for the single target arm-based reaching task. The horizontal red dashed line represents human average reaching accuracy in analogous reaching tasks \citep[based on][]{van2004role}.
\textbf{c}. Accuracy comparison across algorithms for the multi-target arm-based reaching task. \textbf{d}-\textbf{f}. Accuracy comparison between MB-DPG and the Model-based RL algorithm under three different levels of model biases, \textbf{d}. low, \textbf{e}. medium and \textbf{f}. high in the multi-target reaching task.}
\label{fig:Arm_diagram}
\end{figure}

In particular, we introduced errors in the model using a bias in the estimation of sensory outcomes (i.e. the targets learned by the model are biased away from the true outcomes).
We tested under three levels of bias (low, medium and high; see Methods \ref{modelLearning}). 
In figure (\ref{fig:Arm_diagram}d), we can see that under low model bias, both algorithms perform well (e.g. similar performance to no bias condition - see figure \ref{fig:Arm_diagram}c). 
Nevertheless, as the model bias increases (i.e. figure (\ref{fig:Arm_diagram}e,f), the model-based algorithm becomes greatly impaired, while MB-DPG roughly maintains the same level of performance (i.e. the optimum relative to no bias). 
These results show that MB-DPG is considerably more robust to model biases than model-based learning, due to its human-like property of observed-outcome dependent learning. 
These results can be understood using the schematic in Fig.~\ref{fig:Learning_diagram}.
In particular, in MB-DPG, we use the gradient of the reward, $dr/dy$, for the true outcome.
Intuitively, this is an arrow pointing from the actual outcome to the target, and we change the actions such that the outcome in the next trial moves in that direction (Fig.~\ref{fig:Learning_diagram}d).
Thus, as long as $dy/da$ has the correct sign, MB-DPG should converge to the right answer.
However, in model-based learning we use the gradient of the reward, $dr/dy$, for the estimated outcome.
Intuitively, this is an arrow pointing from the estimated outcome to the target, again we change the actions such that the outcome on the next trial moves in that direction.
If the model is wrong, then $\hat{y}$ can be very far from the true outcome, meaning that the ``arrow'' formed by $dr/dy$ is also is in the wrong direction (as shown in Fig.~\ref{fig:Learning_diagram}c).  
In effect, model-based is trying to hit a target biased by model error.
This provides a potential explanation for why human may have evolved to use an approach more similar to MB-DPG (error-based learning) rather than pure model-based approaches.

\section{Discussion}

Deep RL methods have proven successful in complex simulated and robotic ``motor control'' tasks.
We took some of the first steps towards investigating whether these deep RL methods can help us understand human motor learning.
In particular, we began by asking whether any of the major families of modern deep RL algorithm resembles a key motor learning framework in neuroscience, error-based learning.
Using a task with mirror-reversal of visual feedback, we found that canonical deep RL algorithms do not reproduce human behaviour, which is usually accounted for in terms of error-based learning.
This finding implies key differences between human motor learning and modern deep RL.
This is perhaps most surprising in the case of model-based RL, because model based RL, like error-based learning, uses gradients of sensory outcomes with respect to actions to drive updates to a policy.
However, model-based RL relies on the \textit{predicted} (sensory) outcome of actions to drive learning (e.g.\ predicting a reach with a leftward error leads to a rightward correction on the next attempt).
In contrast, (human) error-based learning, is driven by the \textit{observed} sensory outcome of actions (e.g. observing a reach with a leftward error leads to a rightward correction on the next attempt), explaining the worsening of performance documented under mirror-reversal \citep[e.g.][]{hadjiosif2021did,lillicrap2013adapting,wang2021implicit,krakauer2019motor,kasuga2015alteration}. 
Additionally, despite superficial similarities, we showed model-free RL in the form of both policy-gradient (REINFORCE) and value-based (DDPG) approaches differ from (human) error-based learning because these methods completely disregard the sensory outcome of actions, relying only on action-reward pairings.
To bridge this fundamental difference between canonical RL algorithms and (human) error-based learning, we proposed MB-DPG, a (deep) RL algorithm that explicitly relies on \textit{observed} rather than \textit{predicted} sensory outcomes.
We showed MB-DPG captures human performance under mirror-reversal and rotational perturbations.
Next, we considered using these methods to learn a reaching task from scratch, and found that the performance of model-based RL and MB-DPG was roughly similar and better than model-free approaches.
However, we found that MB-DPG was considerably more robust to model misspecification, e.g. caused by tiredness or injury, than model-based approaches.



We believe these last two results are critical for understanding why humans appear to use an error-based strategy such as MB-DPG for motor learning.
In particular, rapid learning of motor behaviours such as effective flight from predators or effective hunting is critical for animal survival.
This would suggest either MB-DPG or model-based RL would be preferable, as we showed that they learn faster than model-free approaches.
However, rapid learning is not the only challenge faced by humans and animals.
Additionally, they face rapid changes in the environment (e.g.\ wind speed and direction) and bodily state (e.g.\ through tiredness or injury), and these changes imply that the forward model will become mis-specified \citep{wolpert2001perspectives,shadmehr2010error}. 
This would suggest that error-based learning would be preferable, as we showed that error-based learning in the form of MB-DPG is more robust to mis-specification than model-based learning.


Our work also has important implications for deep RL.
In particular, it is widely recognised that human and animal motor control and learning are far superior to that of the current generation of robotic deep RL systems \citep[][]{merel2019hierarchical}.
This raises an important question: can we take inspiration from human motor systems to build better deep RL systems?
We sought to take a step along this path, by attempting to understand how human motor control relates to canonical reinforcement learning algorithms.
Indeed, we found that there were key qualitative differences.
We therefore introduced a new method, MB-DPG which combined human error-based strategies for motor adaptation with deep neural networks and other techniques from deep RL.
In particular, we showed that by taking inspiration from (human) error-based learning, we can achieve a model-based RL algorithm, MB-DPG, which mitigates the negative effects of biased forward models (i.e. model mis-specification).
This property addresses a well-know problem in the RL literature \citep[][]{deisenroth2010reducing, deisenroth2011pilco}. 
However, there are still key challenges to realising the full potential of MB-DPG and (human) error-based learning within RL settings.
Perhaps the biggest challenge is that when backpropagating through chaotic physical dynamics across long time windows, often we find gradients explode and thus become uninformative \citep[][]{xu2022accelerated, clavera2020model}.
This suggests that humans use complex adaptive strategies for mitigating gradient explosion in these settings, perhaps backpropagating only through short time windows corresponding to a single action \citep[e.g. similarly to ][]{xu2022accelerated, clavera2020model}.
However, it is not yet clear how humans or deep RL algorithms might adaptively learn to segment long, complex action sequences into their individual component actions, and even whether the notion of individual component actions can be formally defined in long, complex, continuous motor sequences.

Moveover, it is still unclear whether MB-DPG can be directly implemented in neural hardware.
In particular, our current implementation of MB-DPG relies on backpropagation through a learned forward model to estimate the necessary gradients \citep[i.e. relying on the approach proposed by][]{jordan1992forward}.
There are now several proposals for how gradients might be backpropagated in the brain \citep[][]{lillicrap2020backpropagation,whittington2019theories,sacramento2018dendritic}, though it is not yet clear whether the brain actually uses any of these mechanisms.
Alternatively, as MB-DPG evaluates the errors based on the observed, rather than predicted, outcomes, there is the possibility that the required gradients could be learned directly from experience instead of being computed by backpropagating through the forward-model.
In that case, we might expect humans to take exploratory actions strategies that would be informative about these gradients.
For instance humans might on sequential trials make closely related actions (or actions that only vary one component) in order to try to encode information about gradients (i.e.\ how changes to the action change the outcome) \citep[e.g.][]{wu2014temporal}.
Superficially, it seems plausible humans learn better estimates of relative changes in sensory outcomes (i.e. gradients) than exact estimates of sensory outcomes (e.g. it feels much easier to predict that a stronger kick to a ball will make it travel further, than predicting its exact location).
A thorough investigation of these issues involves understanding both behaviour and microcircuits, and constitutes an important avenue for future research.

Overall, our work highlights the gap between current deep RL methods and human and animal motor learning and offers a route to closing this gap.
We hope this will not only present the possibility of developing future deep RL methods which obtain improved performance by mirroring human motor learning, but also present the possibility of leveraging deep RL to understand the full depth and complexity of human and animal motor behaviours.

\section{Methods}

\subsection{Canonical deep RL algorithms}
\label{algorithms}

We employ the three major families of policy-gradient deep RL algorithms: REINFORCE (i.e. a canonical policy gradient method), Deep Deterministic Policy Gradient (i.e. a value-based method) and a policy-gradient model-based algorithm.
Ultimately, all algorithms estimate gradients to find the policy parameters, $\phi$, that maximise the reward, $r$, but each algorithm uses a different approach to estimating the gradients, $dr/d\phi$. 
For all algorithms (including MB-DPG), the action, $a$, was generated from a Gaussian policy, $\pi_\phi(a)$, with mean parametrised by a function of the target value, $y^*$, denoted as $f_{\phi}(y^*)$:   
\begin{align}\label{gaussianPolicy}
a_t &\sim \pi_{\phi}(a) =  \mathcal{N}(f_{\phi}(y^*) \ ,  \sigma) 
\end{align} 
where $\sigma$ represents the standard deviation of the Gaussian policy, which was fixed and treated as a hyper-parameter.
This Gaussian policy ensured sufficient exploration.
The mean function, $f_{\phi}()$, was linear for the mirror-reversal task and was a neural network for the arm-reaching task.

Note that we consider ballistic movement tasks, which in effect have one ``action'', chosen before movement onset, and one reward as an outcome.
In the formal RL setting, this corresponds to a bandit task.

First, we consider REINFORCE, a stochastic policy gradient method \citep{williams1992simple}. 
REINFORCE estimates the gradient of the reward using,
\begin{align}
  \nabla_\phi r &= \E_{a \sim \pi_\phi}\sqb{r \nabla_\phi \log \pi_\phi(a)}.
\end{align}
where $r$ is the reward resulting from taking action $a$.


Second we considered deep deterministic policy gradient (DDPG) \citep{lillicrap2015continuous, silver2014deterministic}. 
In DDPG, we start by learning a differentiable, neural network proxy for the true reward, $Q_\theta(a)$ with parameters $\theta$. 
We use the mean-squared error to the true reward as a loss,
\begin{align}
  \mathcal{L}_\text{Q} &= (r - Q_\theta(a))^2.
\end{align}
Then we estimate the true gradient of the reward using this differentiable proxy,
\begin{align}
  \nabla_\phi r &=  \nabla_\phi Q_\theta(a)\Bigr\rvert_{a = f_{\phi}(y^*)}
\end{align}
For the mirror reversal task, we used a quadratic $Q_\theta(a)$ and for the reaching tasks, we used a neural network.
%
%
%

Third, we considered a policy gradient model-based algorithm \citep[e.g.][]{clavera2020model,heess2016learning}. 
In this approach, the gradient of the reward relative to the action parameters is computed by back-propagating the reward through a (learned) model of the dynamics. 
This results in the following gradient estimate:
\begin{align}\label{model_grad}
  \frac{dr}{d\phi} &\approx \frac{dr}{d \hat{y}} \frac{d \hat{y}}{da} \frac{da}{d\phi} 
\end{align}
where $\frac{d\yh}{da}$ represents the action-to-outcome relation, computed by differentiating through the learnt model of the dynamics. 
Note how in this model-based approach, the directed error signal (i.e.\ $\frac{dr}{dy}$) is evaluated relative to the model prediction, $\hat{y}$, rather than the observed outcome of actions, $y$ (i.e.\ $\frac{dr}{dy} \ = \frac{dr}{d \hat{y}}$). 
This approach is equivalent to using a learned model to predict the outcome of an action, computing the reward based on that predicted outcome, and backpropagating the reward to the action parameters through the learnt model.
Note that a memory buffer was used to train the critic function of DDPG as well as the learned model of the dynamics for both Model-based and MB-DPG algorithms. 

\subsection{Learning the model of the dynamics}
\label{modelLearning}

A model of the dynamics had to be learnt for both the model-based and MB-DPG algorithms to estimate the action-to-outcome relationship. 
A model of the dynamics, denoted as $\hat{y}$, can be learnt through experience, by minimising the following loss across $T$ trials:
\begin{align}\label{modelLoss}
\mathcal{L}_\text{M} &= \sum_{t=1}^T  (\hat{y}(a_t) - y(a_t))^2
\end{align}
Here $y(a_t)$ denotes the observed outcome of the action taken on the $t$ trial. 
In the mirror-reversal task, $y$ consisted of a scalar value, while in the arm-based reaching tasks, it consisted of a 4 dimensional vector, representing the two joint angles and angle velocities. 
Conversely, $\hat{y}(a_t)$ denotes the model prediction of taking action $a_t$. 
The model $\hat{y}()$ consisted of a linear function and neural network for the mirror-reversal paradigm and the arm-based reaching task respectively.

To induce a variable degree of bias in the model of the dynamics, we added a fixed (random) offset, denoted as $b$, to the observed outcome, $y_{obs}$ when training the model:
\begin{align}\label{modelLossBiased}
\mathcal{L}_{\text{BM}} &= \sum_{t=1}^T  (\hat{y}(a_t) - (y(a_t) + b))^2
\end{align}
where $\mathcal{L}_{BM}$ denotes the modified  model training loss across $T$ trials (i.e. biased), and:
\begin{align}\label{modelLossBiased}
b &\sim \mathcal{N}(0 \ ,  \sigma_b)
\end{align}
where $\sigma_b$ controlled the amount of bias (offset) induced in the model of the dynamics. 
We tested for three different sizes of $\sigma_b$, corresponding to low, medium and high bias induction (i.e. $\sigma_b = 0.05; 0.15; 0.25$ respectively). 
Note that $b$ does not depend on $t$, it is sampled once before each training procedure and left constant for all trials, so that to induce a (fixed) bias in the model of the dynamics. 
This analysis was carried out in the more complex arm-based multi-target reaching task to test model-based and MB-DPG robustness to model biases under more realistic settings.

\subsection{Mirror-reversal task}

To test the mirror-reversal paradigm, we chose a simplified environment, following a similar approach to \citet{hadjiosif2021did}. 
This environment was simple enough for us to be able to analyse how each algorithm behaved under mirror-reversal. 
The observed outcome, $y$, was related to the action, $a$ by the following linear model:
\begin{align}\label{mirrorModel}
y &= wa +c 
\end{align} 
The goal of the task was to find the action, $a$, that maximises the reward over a single step. 
For each trial, the reward was computed as the negative value of the squared distance between the observed action outcome, $y$, and the position of the target, denoted as $y^*$:
\begin{align}\label{rwd_func}
  r &= -(y^* - y)^2.
\end{align} 
So the maximum reward was achieved at the target (i.e. when the action achieved the desired target, $y^*$). 
The mirror-reversal of visual feedback, was induced by flipping the sign of $w$ in the environment (equation \ref{mirrorModel}). 
Since we set the the target, $y^*$, and the environment intercept, $c$ in equation (\ref{mirrorModel}), to the same value (i.e. $c = y^*$), we ensured the mirror-reflection occurred along the target direction.
However, this is not a requirement and we did it in this way to replicate \citet[][]{hadjiosif2021did} 's settings.
Consequently, the ``best'' action which hits the target was left uncharged by the mirror-reversal, while all the other actions led to the mirror-reflected outcome.
The mirror-reversal was induced when each algorithm reached a good level of accuracy on the task (i.e. accuracy <0.1), as with the human participants in \citet{hadjiosif2021did}.

\subsection{Rotation task}

For the rotation task, we used the same set-up as in the mirror-reversal task.
The rotational perturbation was induced by adding a fixed bias to the true outcomes,
\begin{align}\label{rotationModel}
y &= wa +c + \theta
\end{align}
where $\theta$ represents the amount of rotational perturbation applied.
In this way, all action outcomes, $y$, were displaced (rotated) of $\theta$.
To replicate \citet[][]{krakauer2000learning}, we used a 30 degrees rotation.

\subsection{MB-DPG and human motor learning}
\label{human_MBDPG}

In order to replicate human performance in the mirror-reversal and rotation task (see Fig.~\ref{fig:HumanLike_MBDPG}), we provided MB-DPG with two additional properties: an update saturation mechanism and non-visual sensory feedback (proprioceptive). The update saturation followed a similar approach to \citep[][]{hadjiosif2021did} and was motivated by extensive evidence that human motor learning saturates with the magnitude of the error \citep[][]{krakauer2019motor,marko2012sensitivity,wei2009relevance}.
This update saturation consisted of pushing MB-DPG error update ($\frac{dr}{dy}$) through a saturating tanh function:
\begin{align}\label{saturating_update}
  \frac{dr}{dy} &= \beta \ \tanh{(\frac{1}{\beta} \ \frac{dr}{dy})} 
\end{align}
where $\beta$ is an hyper-parameter controlling the amount of saturation.
Note that this leaves small errors unchanged, as the function is specifically chosen to have gradient $1$ when the error is zero,
\begin{align}
  \left. \frac{d}{dx} \right|_{x=0} \beta \tanh\b{\tfrac{1}{\beta} x} = 1,
\end{align}
but it does saturate larger values for the error, with the maximal values being $\pm \beta$, which appears to be the case for humans \citep[][]{krakauer2019motor,marko2012sensitivity,wei2009relevance}.
In addition, non-visual feedback was needed to model human performance under the no-visual feedback trials (see \ref{fig:HumanLike_MBDPG}a, gray shade).
This non-visual feedback was implemented by including information about the undertaken action, $a$ (i.e. priprioceptive), in the reward computation, $r$ (i.e. visual error):
\begin{align}\label{pripriocept_rwd_func}
  r &= -(y^* - y)^2 - \gamma * a^2
\end{align}
where $\gamma$ was an hyper-parameter controlling the weighting on the proprioceptive feedback.

\subsection{Biomechanical model of the human arm}

We employed a simulated biomechanical model of the human arm from the neuroscience literature \citep[][]{berret_chiovetto_nori_pozzo_2011}.
The model consisted of a two-joint rigid body (i.e. shoulder and elbow joints), operating in a 2 dimensional horizontal plane.
This model implies that the position of the hand in space is fully determined by the angles of the two joints.
The upper or lower arm lengths were computed according to standard formulas based on an average person's height of 1.80 meters \citep[see][]{winter2009biomechanics, berret_chiovetto_nori_pozzo_2011}.
The masses of the upper and lower arms were also computed according to standard formulas based on an average person's weight of 80kg.
The biomechanics for this arm model are described by the following dynamical system:

\begin{equation} \label{Arm_system}
 \tau = \mathbf{M}(\theta)\ddot\theta + \mathbf{C}(\theta,\dot\theta)\dot\theta  + \mathbf{F} \dot\theta 
 \end{equation}
 
 \begin{equation} \label{Arm_system2}
 \ddot\tau = a(t)
 \end{equation}
 
\noindent where the quantities $ \mathbf{M} , \mathbf{C} , \mathbf{F}$ respectively represent the inertia, Coriolis and viscosity matrices and were set to realistic values based on \citep[][]{winter2009biomechanics, berret_chiovetto_nori_pozzo_2011}.
Conversely, $ \theta = (\theta_1 , \theta_2)^\intercal $ represents the joint angle vector (i.e. two joints), $ \tau = (\tau_1, \tau_2)^\intercal$ represents the joint torque vector and $a(t) = (a_1(t), a_2(t)$) indicates the vector-valued function of the actions as a function of time. 
For simplicity, the two action values at time $t$, $(a_1(t), a_2(t))$, respectively controlled the acceleration of the shoulder and elbow torques,$(\ddot\tau_1, \ddot\tau_2)$ in a mechanical manner (i.e. noise free).
Each simulated arm trajectory lasted for 0.4 seconds, divided into 100 time steps of equal length (i.e. each of length 0.004 seconds) and begun with the arm in the same initial position, namely, with the upper and lower arms respectively laying horizontally and vertically straight (i.e. $\theta = (- \frac{\pi}{2}, \frac{\pi}{2})^\intercal$ - similarly to human reaching experiments onto a 2d planar surface) (see Fig. \ref{fig:Arm_diagram}a). 
Each arm trajectory was controlled by a two hundred dimensional action vector given by the policy (i.e.\ one action for each of the two joint torque acceleration for 100 time steps). 
It is worth noticing each entry in this action vector depended on time, but not on the state of the arm at that time point. 
This is because the model aimed to simulate a ballistic (feedforward) reach.
The dynamics of the system, (\ref{Arm_system}) and (\ref{Arm_system2}) , were approximated with a 4th order Runge-Kutta method, since the system could not be integrated analytically.

The reward was computed as the negative of the euclidean distance between the final index finger position and the target locations, plus a penalty on the arm end-velocity in euclidean coordinates. The latter ensured the arm reached the target with a velocity close to zero.

\begin{align}\label{ArmBased_rwd_func}
  r &= -(y^* - y)^2 + \beta  \dot y
\end{align}

\noindent where $y^*$ and $y$ respectively denote the target location and the (observed) index finger location after a reaching attempt, while $\dot y$ represents the arm final velocity. The hyper-parameter $\beta$ controlled the penalty on the end point velocity.

\subsection{Hyper-parameter tuning}
\label{hyperSearch}

In order to accurately compare the performance of Reinforce, DPG, Model-based and MB-DPG, a hyper parameter search was performed for each algorithm. 
Based on this search, the set of hyper-parameters giving the best performance was chosen for each algorithm for the comparison. 
The hyper-parameter search consisted of an exhaustive grid search over a plausible range of values, identified after several trial-error runs. 
This search was carried out across 5 training seeds, which were the same for all 4 algorithms and had randomly been chosen. 
For Reinforce, the grid search was performed over the learning rate of the actor (i.e. 7 values linearly spaced between 0.000005 and 0.001) and the standard deviation of the Gaussian policy (i.e. 7 values linearly spaced between 0.001 and 0.05). 
For DPG, the grid search also included the learning rate of the critic (i.e. 7 values linearly spaced between 0.000001 and 0.0001). 
Finally, for Model-based and MB-DPG, the grid search included the learning rate of the actor (i.e. 7 values linearly spaced between 0.000005 and 0.0005) and of the model (i.e. 7 values linearly spaced between 0.0001 and 0.01), plus the standard deviation of the Gaussian policy (i.e. 7 values linearly spaced between 0.001 and 0.05).
The performance reported in the result section are averaged across 10 randomly chosen test seeds (different from the training ones) for both the single and multi target reaching tasks. This procedure ensures reliability in the results.

\printbibliography 

@article{berret_chiovetto_nori_pozzo_2011, title={Evidence for Composite Cost Functions in Arm Movement Planning: An Inverse Optimal Control Approach}, volume={7}, DOI={10.1371/journal.pcbi.1002183}, number={10}, journal={PLoS Computational Biology}, author={Berret, Bastien and Chiovetto, Enrico and Nori, Francesco and Pozzo, Thierry}, year={2011}}

@article{jordan1992forward,
  title={Forward models: Supervised learning with a distal teacher},
  author={Jordan, Michael I and Rumelhart, David E},
  journal={Cognitive science},
  volume={16},
  number={3},
  pages={307--354},
  year={1992},
  publisher={Elsevier}
}

@article{williams1992simple,
  title={Simple statistical gradient-following algorithms for connectionist reinforcement learning},
  author={Williams, Ronald J},
  journal={Machine learning},
  volume={8},
  number={3},
  pages={229--256},
  year={1992},
  publisher={Springer}
}

@inproceedings{silver2014deterministic,
  title={Deterministic policy gradient algorithms},
  author={Silver, David and Lever, Guy and Heess, Nicolas and Degris, Thomas and Wierstra, Daan and Riedmiller, Martin},
  booktitle={International conference on machine learning},
  pages={387--395},
  year={2014},
  organization={PMLR}
}

@article{lillicrap2015continuous,
  title={Continuous control with deep reinforcement learning},
  author={Lillicrap, Timothy P and Hunt, Jonathan J and Pritzel, Alexander and Heess, Nicolas and Erez, Tom and Tassa, Yuval and Silver, David and Wierstra, Daan},
  journal={arXiv preprint arXiv:1509.02971},
  year={2015}
}

@article{hassabis2017neuroscience,
  title={Neuroscience-inspired artificial intelligence},
  author={Hassabis, Demis and Kumaran, Dharshan and Summerfield, Christopher and Botvinick, Matthew},
  journal={Neuron},
  volume={95},
  number={2},
  pages={245--258},
  year={2017},
  publisher={Elsevier}
}

@incollection{sutton1990time,
  title={Time-derivative models of Pavlovian reinforcement.},
  editor={M. Gabriel, and J. Moore},
  booktitle={Learning and computational neuroscience: Foundations of adaptive networks},
  author={Sutton, Richard S and Barto, Andrew G},
  year={1990},
  pages={497–537},
  publisher={The MIT Press}
}

@article{schultz1997neural,
  title={A neural substrate of prediction and reward},
  author={Schultz, Wolfram and Dayan, Peter and Montague, P Read},
  journal={Science},
  volume={275},
  number={5306},
  pages={1593--1599},
  year={1997},
  publisher={American Association for the Advancement of Science}
}

@article{montague1996framework,
  title={A framework for mesencephalic dopamine systems based on predictive Hebbian learning},
  author={Montague, P Read and Dayan, Peter and Sejnowski, Terrence J},
  journal={Journal of neuroscience},
  volume={16},
  number={5},
  pages={1936--1947},
  year={1996},
  publisher={Soc Neuroscience}
}

@article{niv2009reinforcement,
  title={Reinforcement learning in the brain},
  author={Niv, Yael},
  journal={Journal of Mathematical Psychology},
  volume={53},
  number={3},
  pages={139--154},
  year={2009},
  publisher={Elsevier}
}

@inproceedings{barto1995adaptive,
  title={Adaptive critics and the basal ganglia},
  author={Barto, Andrew G},
  booktitle={Models of Information Processing in the Basal Ganglia},
  editor={J. C. Houk, and J. Davis, and D. Beiser},
  pages={215-232},
  publisher={Cambridge, MA: MIT Press},
  year={1995}
}

@inproceedings{sutton1987temporal,
  title={A temporal-difference model of classical conditioning},
  author={Sutton, Richard S and Barto, Andrew G},
  booktitle={Proceedings of the ninth annual conference of the cognitive science society},
  pages={355--378},
  year={1987},
  organization={Seattle, WA}
}

@article{bayer2005midbrain,
  title={Midbrain dopamine neurons encode a quantitative reward prediction error signal},
  author={Bayer, Hannah M and Glimcher, Paul W},
  journal={Neuron},
  volume={47},
  number={1},
  pages={129--141},
  year={2005},
  publisher={Elsevier}
}

@article{merel2018hierarchical,
  title={Hierarchical visuomotor control of humanoids},
  author={Merel, Josh and Ahuja, Arun and Pham, Vu and Tunyasuvunakool, Saran and Liu, Siqi and Tirumala, Dhruva and Heess, Nicolas and Wayne, Greg},
  journal={arXiv preprint arXiv:1811.09656},
  year={2018}
}

@article{akkaya2019solving,
  title={Solving rubik's cube with a robot hand},
  author={Akkaya, Ilge and Andrychowicz, Marcin and Chociej, Maciek and Litwin, Mateusz and McGrew, Bob and Petron, Arthur and Paino, Alex and Plappert, Matthias and Powell, Glenn and Ribas, Raphael and others},
  journal={arXiv preprint arXiv:1910.07113},
  year={2019}
}

@article{heess2016learning,
  title={Learning and transfer of modulated locomotor controllers},
  author={Heess, Nicolas and Wayne, Greg and Tassa, Yuval and Lillicrap, Timothy and Riedmiller, Martin and Silver, David},
  journal={arXiv preprint arXiv:1610.05182},
  year={2016}
}

@article{botvinick2020deep,
  title={Deep reinforcement learning and its neuroscientific implications},
  author={Botvinick, Matthew and Wang, Jane X and Dabney, Will and Miller, Kevin J and Kurth-Nelson, Zeb},
  journal={Neuron},
  year={2020},
  volume={107},
  number={4},
  pages={603-616},
  publisher={Elsevier}
}

@article{hadjiosif2021did,
  title={Did we get sensorimotor adaptation wrong? Implicit adaptation as direct policy updating rather than forward-model-based learning},
  author={Hadjiosif, Alkis M and Krakauer, John W and Haith, Adrian M},
  journal={Journal of Neuroscience},
  volume={41},
  number={12},
  pages={2747--2761},
  year={2021},
  publisher={Soc Neuroscience}
}

@article{abdelghani2008sensitivity,
  title={Sensitivity derivatives for flexible sensorimotor learning},
  author={Abdelghani, M N and Lillicrap, Timothy P and Tweed, Douglas B},
  journal={Neural computation},
  volume={20},
  number={8},
  pages={2085--2111},
  year={2008},
  publisher={MIT Press One Rogers Street, Cambridge, MA 02142-1209, USA journals-info~…}
}

@article{izawa2011learning,
  title={Learning from sensory and reward prediction errors during motor adaptation},
  author={Izawa, Jun and Shadmehr, Reza},
  journal={PLoS computational biology},
  volume={7},
  number={3},
  pages={e1002012},
  year={2011},
  publisher={Public Library of Science San Francisco, USA}
}

@article{tseng2007sensory,
  title={Sensory prediction errors drive cerebellum-dependent adaptation of reaching},
  author={Tseng, Ya-weng and Diedrichsen, Jörn and Krakauer, John W and Shadmehr, Reza and Bastian, Amy J},
  journal={Journal of neurophysiology},
  volume={98},
  number={1},
  pages={54--62},
  year={2007},
  publisher={American Physiological Society}
}

@inproceedings{deisenroth2011pilco,
  title={PILCO: A model-based and data-efficient approach to policy search},
  author={Deisenroth, Marc P and Rasmussen, Carl E},
  booktitle={Proceedings of the 28th International Conference on machine learning (ICML-11)},
  pages={465--472},
  year={2011},
  organization={Citeseer}
}

@article{deisenroth2010reducing,
  title={Reducing model bias in reinforcement learning},
  author={Deisenroth, Marc P and Rasmussen, Carl E},
  year={2010},
  publisher={Citeseer}
}

@article{wolpert2011principles,
  title={Principles of sensorimotor learning},
  author={Wolpert, Daniel M and Diedrichsen, J{\"o}rn and Flanagan, J Randall},
  journal={Nature Reviews Neuroscience},
  volume={12},
  number={12},
  pages={739--751},
  year={2011},
  publisher={Nature Publishing Group}
}

@article{hosp2011dopaminergic,
  title={Dopaminergic projections from midbrain to primary motor cortex mediate motor skill learning},
  author={Hosp, Jonas A and Pekanovic, Ana and Rioult-Pedotti, Mengia S and Luft, Andreas R},
  journal={Journal of Neuroscience},
  volume={31},
  number={7},
  pages={2481--2487},
  year={2011},
  publisher={Soc Neuroscience}
}

@article{luft2009dopaminergic,
  title={Dopaminergic signals in primary motor cortex},
  author={Luft, Andreas R and Schwarz, Stefanie},
  journal={International Journal of Developmental Neuroscience},
  volume={27},
  number={5},
  pages={415--421},
  year={2009},
  publisher={Elsevier}
}

@article{synofzik2006internalizing,
  title={Internalizing agency of self-action: perception of one's own hand movements depends on an adaptable prediction about the sensory action outcome},
  author={Synofzik, Matthis and Thier, Peter and Lindner, Axel},
  journal={Journal of neurophysiology},
  volume={96},
  number={3},
  pages={1592--1601},
  year={2006},
  publisher={American Physiological Society}
}

@article{shadmehr2010error,
  title={Error correction, sensory prediction, and adaptation in motor control},
  author={Shadmehr, Reza and Smith, Maurice A and Krakauer, John W},
  journal={Annual review of neuroscience},
  volume={33},
  pages={89--108},
  year={2010},
  publisher={Annual Reviews}
}

@article{mazzoni2006implicit,
  title={An implicit plan overrides an explicit strategy during visuomotor adaptation},
  author={Mazzoni, Pietro and Krakauer, John W},
  journal={Journal of neuroscience},
  volume={26},
  number={14},
  pages={3642--3645},
  year={2006},
  publisher={Soc Neuroscience}
}

@article{kitazawa1998cerebellar,
  title={Cerebellar complex spikes encode both destinations and errors in arm movements},
  author={Kitazawa, Shigeru and Kimura, Tatsuya and Yin, Ping-Bo},
  journal={Nature},
  volume={392},
  number={6675},
  pages={494--497},
  year={1998},
  publisher={Nature Publishing Group}
}

@article{ito2013error,
  title={Error detection and representation in the olivo-cerebellar system},
  author={Ito, Masao},
  journal={Frontiers in neural circuits},
  volume={7},
  pages={1},
  year={2013},
  publisher={Frontiers}
}

@article{dean2010cerebellar,
  title={The cerebellar microcircuit as an adaptive filter: experimental and computational evidence},
  author={Dean, Paul and Porrill, John and Ekerot, Carl-Fredrik and J{\"o}rntell, Henrik},
  journal={Nature Reviews Neuroscience},
  volume={11},
  number={1},
  pages={30--43},
  year={2010},
  publisher={Nature Publishing Group}
}

@article{miall1996forward,
  title={Forward models for physiological motor control},
  author={Miall, R Chris and Wolpert, Daniel M},
  journal={Neural networks},
  volume={9},
  number={8},
  pages={1265--1279},
  year={1996},
  publisher={Elsevier}
}

@article{lillicrap2013adapting,
  title={Adapting to inversion of the visual field: a new twist on an old problem},
  author={Lillicrap, Timothy P and Moreno-Brise{\~n}o, Pablo and Diaz, Rosalinda and Tweed, Douglas B and Troje, Nikolaus F and Fernandez-Ruiz, Juan},
  journal={Experimental brain research},
  volume={228},
  number={3},
  pages={327--339},
  year={2013},
  publisher={Springer}
}

@article{wang2021implicit,
  title={Implicit adaptation to mirror-reversal is in the correct coordinate system but the wrong direction},
  author={Wang, Tianhe and Taylor, Jordan},
  journal={bioRxiv},
  year={2021},
  publisher={Cold Spring Harbor Laboratory}
}

@book{winter2009biomechanics,
  title={Biomechanics and motor control of human movement},
  author={Winter, David A},
  year={2009},
  publisher={John Wiley \& Sons}
}

@article{schulman2017proximal,
  title={Proximal policy optimization algorithms},
  author={Schulman, John and Wolski, Filip and Dhariwal, Prafulla and Radford, Alec and Klimov, Oleg},
  journal={arXiv preprint arXiv:1707.06347},
  year={2017}
}

@article{matheron2019problem,
  title={The problem with DDPG: understanding failures in deterministic environments with sparse rewards},
  author={Matheron, Guillaume and Perrin, Nicolas and Sigaud, Olivier},
  journal={arXiv preprint arXiv:1911.11679},
  year={2019}
}

@article{van2004role,
  title={The role of execution noise in movement variability},
  author={Van Beers, Robert J and Haggard, Patrick and Wolpert, Daniel M},
  journal={Journal of neurophysiology},
  volume={91},
  number={2},
  pages={1050--1063},
  year={2004},
  publisher={American Physiological Society}
}

@article{krakauer2019motor,
  title={Motor learning},
  author={Krakauer, John W and Hadjiosif, Alkis M and Xu, Jing and Wong, Aaron L and Haith, Adrian M},
  journal={Compr Physiol},
  volume={9},
  number={2},
  pages={613--663},
  year={2019}
}

@article{wolpert2001perspectives,
  title={Perspectives and problems in motor learning},
  author={Wolpert, Daniel M and Ghahramani, Zoubin and Flanagan, J Randall},
  journal={Trends in cognitive sciences},
  volume={5},
  number={11},
  pages={487--494},
  year={2001},
  publisher={Elsevier}
}

@article{kasuga2015alteration,
  title={Alteration of a motor learning rule under mirror-reversal transformation does not depend on the amplitude of visual error},
  author={Kasuga, Shoko and Kurata, Makiko and Liu, Meigen and Ushiba, Junichi},
  journal={Neuroscience research},
  volume={94},
  pages={62--69},
  year={2015},
  publisher={Elsevier}
}

@article{krakauer2000learning,
  title={Learning of visuomotor transformations for vectorial planning of reaching trajectories},
  author={Krakauer, John W and Pine, Zachary M and Ghilardi, Maria-Felice and Ghez, Claude},
  journal={Journal of neuroscience},
  volume={20},
  number={23},
  pages={8916--8924},
  year={2000},
  publisher={Soc Neuroscience}
}

@article{marko2012sensitivity,
  title={Sensitivity to prediction error in reach adaptation},
  author={Marko, Mollie K and Haith, Adrian M and Harran, Michelle D and Shadmehr, Reza},
  journal={Journal of neurophysiology},
  volume={108},
  number={6},
  pages={1752--1763},
  year={2012},
  publisher={American Physiological Society Bethesda, MD}
}

@article{wei2009relevance,
  title={Relevance of error: what drives motor adaptation?},
  author={Wei, Kunlin and Kording, Konrad},
  journal={Journal of neurophysiology},
  volume={101},
  number={2},
  pages={655--664},
  year={2009},
  publisher={American Physiological Society}
}

@misc{Rohatgi2020,
  url = {https://automeris.io/WebPlotDigitizer},
  author = {Rohatgi,  Ankit},
  title = {Webplotdigitizer: Version 4.5},
  year = {2021}
}

@article{rieser1995calibration,
  title={Calibration of human locomotion and models of perceptual-motor organization.},
  author={Rieser, John J and Pick, Herbert L and Ashmead, Daniel H and Garing, Anne E},
  journal={Journal of Experimental Psychology: Human Perception and Performance},
  volume={21},
  number={3},
  pages={480},
  year={1995},
  publisher={American Psychological Association}
}

@article{merel2019hierarchical,
  title={Hierarchical motor control in mammals and machines},
  author={Merel, Josh and Botvinick, Matthew and Wayne, Greg},
  journal={Nature communications},
  volume={10},
  number={1},
  pages={1--12},
  year={2019},
  publisher={Nature Publishing Group}
}

@article{clavera2020model,
  title={Model-augmented actor-critic: Backpropagating through paths},
  author={Clavera, Ignasi and Fu, Violet and Abbeel, Pieter},
  journal={arXiv preprint arXiv:2005.08068},
  year={2020}
  }

@article{wu2014temporal,
  title={Temporal structure of motor variability is dynamically regulated and predicts motor learning ability},
  author={Wu, Howard G and Miyamoto, Yohsuke R and Castro, Luis Nicolas Gonzalez and {\"O}lveczky, Bence P and Smith, Maurice A},
  journal={Nature neuroscience},
  volume={17},
  number={2},
  pages={312--321},
  year={2014},
  publisher={Nature Publishing Group}
}

@article{watkins1992q,
  title={Q-learning},
  author={Watkins, Christopher JCH and Dayan, Peter},
  journal={Machine learning},
  volume={8},
  number={3},
  pages={279--292},
  year={1992},
  publisher={Springer}
}

@article{lillicrap2020backpropagation,
  title={Backpropagation and the brain},
  author={Lillicrap, Timothy P and Santoro, Adam and Marris, Luke and Akerman, Colin J and Hinton, Geoffrey},
  journal={Nature Reviews Neuroscience},
  volume={21},
  number={6},
  pages={335--346},
  year={2020},
  publisher={Nature Publishing Group}
}

@article{whittington2019theories,
  title={Theories of error back-propagation in the brain},
  author={Whittington, James CR and Bogacz, Rafal},
  journal={Trends in cognitive sciences},
  volume={23},
  number={3},
  pages={235--250},
  year={2019},
  publisher={Elsevier}
}

@article{sacramento2018dendritic,
  title={Dendritic cortical microcircuits approximate the backpropagation algorithm},
  author={Sacramento, Jo{\~a}o and Ponte Costa, Rui and Bengio, Yoshua and Senn, Walter},
  journal={Advances in neural information processing systems},
  volume={31},
  year={2018}
}

@article{xu2022accelerated,
  title={Accelerated policy learning with parallel differentiable simulation},
  author={Xu, Jie and Makoviychuk, Viktor and Narang, Yashraj and Ramos, Fabio and Matusik, Wojciech and Garg, Animesh and Macklin, Miles},
  journal={arXiv preprint arXiv:2204.07137},
  year={2022}
}

@article{fischer2021reinforcement,
  title={Reinforcement learning control of a biomechanical model of the upper extremity},
  author={Fischer, Florian and Bachinski, Miroslav and Klar, Markus and Fleig, Arthur and M{\"u}ller, J{\"o}rg},
  journal={Scientific Reports},
  volume={11},
  number={1},
  pages={1--15},
  year={2021},
  publisher={Nature Publishing Group}
}

\end{document}